\documentclass[final]{elsart5p}
\pdfoutput=1
\usepackage{natbib}
\usepackage{graphicx}
\usepackage{amssymb}
\usepackage{longtable}
\usepackage{lscape}

\newcommand\afrho{\mbox{$Af\rho$}}
\newcommand\micron{\mbox{$\mu$m}}%
\newcommand\arcdeg{\mbox{$^\circ$}}%
\newcommand\arcsec{\mbox{$^{\prime\prime}$}}
\newcommand\degr{\arcdeg}%

\newcommand\wcmumsr{W~cm$^{-2}$~\micron$^{-1}$~sr$^{-1}$}
\newcommand\wcmum{W~cm$^{-2}$~\micron$^{-1}$}
\newcommand\sun{\odot}%
\newcommand\inv{~$^{-1}$}
\newcommand\kms{km~s$^{-1}$}

\newcommand\coo{CO$_2$}
\newcommand\water{H$_2$O}
\newcommand\sst{\textit{Spitzer Space Telescope}}
\newcommand\spitzer{\textit{Spitzer}}

\newcommand\di{\textit{Deep Impact}}
\newcommand\stardust{\textit{Stardust}}

\newcommand\iso{\textit{ISO}}

\begin{document}

\begin{frontmatter}

\title{The Composition of Dust in Jupiter-Family Comets Inferred From
  Infrared Spectroscopy}

\author[a]{Michael S. Kelley\corauthref{cor}} and
\corauth[cor]{Corresponding author.}
\ead{msk@physics.ucf.edu}
\author[b]{Diane H. Wooden}
\ead{Diane.H.Wooden@nasa.gov}

\address[a]{Department of Physics, University of Central Florida, 4000
  Central Florida Blvd., Orlando, FL 32816-2385, USA}
\address[b]{NASA Ames Research Center, Space Science Division, MS
  245-1, Moffett Field, CA 94035-1000, USA}

\begin{abstract}

We review the composition of Jupiter-family comet dust as inferred
from infrared spectroscopy.  We find that Jupiter-family comets have
10~\micron{} silicate emission features with fluxes roughly 20--25\%
over the dust continuum (emission strength 1.20--1.25), similar to the
weakest silicate features in Oort Cloud comets.  We discuss the grain
properties that change the silicate emission feature strength
(composition, size, and structure/shape), and emphasize that thermal
emission from the comet nucleus can have significant influence on the
derived silicate emission strength.  Recent evidence suggests that
porosity is the dominant parameter, although more observations and
models of silicates in Jupiter-family comets are needed to determine
if a consistent set of grain parameters can explain their weak
silicate emission features.  Models of 8~m telescope and
\textit{Spitzer Space Telescope} observations have shown that
Jupiter-family comets have crystalline silicates with abundances
similar to or less than those found in Oort Cloud comets, although the
crystalline silicate mineralogy of comets 9P/Tempel and C/1995~O1
(Hale-Bopp) differ from each other in Mg and Fe content.  The
heterogeneity of comet nuclei can also be assessed with mid-infrared
spectroscopy, and we review the evidence for heterogeneous dust
properties in the nucleus of comet 9P/Tempel.  Models of dust
formation, mixing in the solar nebula, and comet formation must be
able to explain the observed range of Mg and Fe content and the
heterogeneity of comet 9P/Tempel, although more work is needed in
order to understand to what extent do comets 9P/Tempel and Hale-Bopp
represent comets as a whole.
\end{abstract}


\end{frontmatter}


\section{Introduction}\label{sec:intro}

The dust in Jupiter-family comets (JFCs) is considered to be pristine,
that is, little or no processing of the dust has occurred since
incorporation into the comet nucleus.  We do not equate pristine dust
with pre-solar dust.  The presence of crystalline silicates in the
dust comae of JFCs and Oort Cloud (OC) comets, and the absence of
crystalline silicates in the interstellar medium
\citep[$\leq 2.2$\%;][]{kemper04, kemper05}, strongly suggests that
much of the dust in comets has been thermally processed
($T\gtrapprox1000$~K).  Strong radiogenic heating may occur in comets
of sufficiently large (radius, $R_n$, $>10$--20~km) porous and dust
rich nuclei \citep{prialnik99, prialnik08, merk06, mckinnon08}, but
the presence of highly volatile molecules, such as CO and/or CO$_2$
observed in the comae of 9P/Tempel \citep{ahearn05},
21P/Giacobini-Zinner \citep{mumma00}, and 103P/Hartley
\citep{crovisier00}, suggests that comet interiors formed at very low
temperatures (as low as $\approx30$~K for CO to condense) and have
since remained cold \citep{yamamoto85, crovisier07}.  The simultaneous
presence of low-temperature and high-temperature solar nebula
condensates in comet nuclei is possible if thermally processed dust
can be mixed from the hot inner-solar nebula into the cold comet
forming zone (heliocentric distance, $r_h$, $\gtrapprox 4$~AU) perhaps
by turbulent diffusion \citep{wehrstedt03a, wehrstedt03b, wehrstedt08,
  wooden07} or by large-scale radial (meridional) flows
\citep{keller04, ciesla07, tscharnuter07}.

Mid-infrared (mid-IR) spectroscopy ($\lambda\sim10$~\micron) is an
effective technique for assessing the composition of comet dust in a
large number of comets: 1) the peak thermal emission output from comet
dust at $r_h\approx0.5$--4.0~AU is at $\lambda\approx7-20$~\micron{};
2) the mid-IR covers prominent emission features from silicate
minerals, a major constituent of comet dust; and 3) only low to
moderate spectral resolution ($\lambda/\Delta\lambda\gtrsim100$) is
required to critically assess the composition of comet dust.  By
studying the physical properties of dust ejected by comet nuclei, we
gain insight into the formation mechanism of dust in the solar nebula,
and the structure and dynamics of the proto-planetary disk (PPD).

Through comparisons between dust in the interstellar medium,
interplanetary dust grains \citep{brownlee85}, comet dust, and
asteroidal materials, we may attempt to ascertain the extent to which
PPD processes altered pre-solar dust and mixed it throughout the comet
forming zone \citep{wooden08}.  Being a mixture of pre-solar and solar
nebula processed material, measuring the composition of comet dust
helps us understand both the origin of the constituents of the
primitive solar system, and their subsequent evolution into
planetesimals \citep{ehrendfreund04, wooden08}.  In a similar vein, we
can use JFCs to learn about the compositions of Scattered Disk objects
since these bodies are dynamically linked \citep{duncan04} and
therefore their dust properties may be related.  In contrast, Kuiper
Belt objects from the classical disk likely formed in situ, and
represent a population of icy bodies separate from the Scattered Disk
objects and comet nuclei.  Indeed, Kuiper Belt objects have redder
surface colors, on average, than comet nuclei \citep{jewitt02},
indicating that they have a different origin or evolutionary history.

Since mid-infrared spectroscopy is effective at measuring comet dust
properties, it offers important insight into the structure of the PPD.
We can use mid-IR observations of comet dust to explore the
heterogeneity of the PPD on sub-kilometer and super-kilometer (i.e.,
possibly tens of AU) scales by measuring the heterogeneity of comet
nuclei.  Dust may be ejected into the comet coma from a localized
region on the nucleus at different epochs either by natural or
unnatural processes, e.g., seasonal variations in insolation,
outbursts, or collisions (\di{}-like events or natural).  If the dust
composition of the ambient coma differs from the localized regions,
then we may conclude that the comet nucleus is heterogeneous.  Nucleus
heterogeneity could arise from the aggregation of varied planetesimals
\citep[see, e.g.,][]{belton07} and evidence for heterogeneity implies
the proto-planetary disk in the comet forming regions was
heterogeneous on kilometer to sub-kilometer sized scales \citep[the
  typical radius of a comet nucleus is of order 1~km;][]{lamy04}.
Alternatively, the dust properties of individual nuclei could be
relatively homogeneous, yet the dust compositions vary from
comet-to-comet.  This second scenario favors PPD heterogeneities on
the super-kilometer scale.  A taxonomy of comets classified according
to their dust properties has the promise to provide evidence
supporting local (spatial or temporal) variations in dust composition
in the comet forming region of the PPD.  Evidence for chemical
heterogeneity on both sub- and super-kilometer scales has been
discovered through observations of comet gas species \citep{dbm04},
e.g., the C$_2$ comet classes \citep{ahearn95}, the organic chemistry
classes \citep{mumma03, dbm04, crovisier07}, the apparent CO
variability of comet 21P/Giacobini-Zinner \citep{mumma00, weaver99},
and the CO$_2$ and H$_2$O composition of comet 9P/Tempel
\citep{feaga07}.

In this paper, we review the composition of Jupiter-family comets as
inferred from infrared spectroscopy: in \S\ref{sec:obs} we review
mid-IR spectroscopic observations of JFCs; in \S\ref{sec:silicates} we
summarize and discuss observations of silicate mineralogy in JFCs, we
demonstrate (\S\ref{sec:nucleus}) that the silicate feature strength
is increased by subtracting emission from the nucleus, but not
increased enough to match the strong silicate features of some OC
comets, and from this perspective we discuss opportunities for
important observations in the future; in \S\ref{sec:tempel} we review
the mid-IR observations and models of comet 9P/Tempel in the context
of the \di{} mission; in \S\ref{sec:9p-discussion} we discuss the
potential significance to our understanding of PPD processes of the
detection of hydrated and Fe-rich minerals identified in comet
9P/Tempel; and, in \S\ref{sec:summary} we summarize the review.

\section{Mid-infrared Spectroscopic Observations}\label{sec:obs}

Mid-infrared spectra, with $\lambda/\Delta\lambda\gtrapprox50$, exist
for 13 Jupiter-family comets and these observations are summarized in
Table~\ref{table:obs}.  We also list whether or not a silicate
emission feature was detected above the continuum for each spectrum,
and the strength of that feature, if available.  It is important to
note that the 10~\micron{} continuum in a mid-IR spectrum of a comet
is composed of thermal emission from dust species that exhibit little
variation in emissivity with wavelength for $\lambda=7-14$~\micron{}.
The common ``continuum'' contributers are amorphous carbon, FeS, and
grains large enough to be optically thick to 10~\micron{} emission
\citep{hanner94, wooden02}.  Thermal emission from the nucleus can
also be an important source of continuum flux, and should be removed
before measuring the silicate emission (see \S\ref{sec:nucleus}).  A
silicate feature has been detected in 8 of the 13 comets in
Table~\ref{table:obs}, and the typical silicate emission band strength
is 1.20--1.25.  The silicate emission band strength is computed from
the ratio of the mean in-band flux between 10.0 and 11.0~\micron{} to
a scaled Planck function that is fit to small sections of the smooth
continuum at $\lambda\leq8.2-8.4$~\micron{} and
$\lambda\geq12.4-12.5$~\micron{} and evaluated at 10.5~\micron{}
\citep{hanner96, sitko04, sugita05}.  Observed silicate band emission
strengths in Oort Cloud comets range from 1.0 to 4.0 \citep{hanner94,
  williams97, harker99aj, sitko04}.  Such strong silicate band
emission has not been observed in a JFC, except in the post-\di{} coma
of 9P/Tempel (\citealt{lisse06}; \citealt{harker07};
\S\ref{sec:tempel}).

The \sst{} \citep{werner04, gehrz07} has significantly increased the
number of spectroscopic observations of JFCs in the mid-IR.  Three
comets with high-quality spectra are listed in Table~\ref{table:obs},
and 9 more are briefly summarized by \citet{kelley06phd}.  This boon
to JFC spectroscopy is largely attributable to the high sensitivities
of the Infrared Spectrograph (IRS) instrument \citep{houck04} and the
high observing efficiency of the \spitzer{} spacecraft.  As future
infrared space telescopes will have larger primary mirror diameters,
such as the JAXA \textit{Space Infrared Telescope for Cosmology and
  Astrophysics} (\textit{SPICA}), and the NASA \textit{James Webb
  Space Telescope}, bright comets may become unaccessible to
space-borne instrumentation.  Instead, ground-based and airborne
telescopes will be used to observe the brightest comets.

Ground-based observations account for 1/2 of the spectra in
Table~\ref{table:obs}; about 2/3 of those spectra were taken at the
NASA's 3~m Infrared Telescope Facility (IRTF).  Eight meter telescopes
hold great promise for the spectroscopic investigation of JFCs.  The
Gemini and Subaru telescopes have already provided three definitive
detections of crystalline silicates in JFCs; the measured
crystalline-to-amorphous sub-\micron{} mass ratios are: $\sim$0.9 for
78P/Gehrels (\citealt{watanabe05}; also discussed in
\citealt{ootsubo07a}), 0.3--0.8 for 9P/Tempel \citep{harker05,
  harker07, sugita05, lisse06}, and qualitatively described as less
than in Hale-Bopp for 73P/Schwassmann-Wachmann \citep{harker06b,
  sitko06irs-iauc}.  Also, Gemini-N and Subaru observations of the
\di{} event provided, in part, the evidence for the heterogeneous
structure and composition of the nucleus of 9P/Tempel \citep{harker07,
  kadono07}.  With the continued use of mid-IR instruments on large
telescopes, and in conjunction with the archive of space-based
observations, we can increase the number of high-quality spectra of
JFCs to statistically significant numbers, which will give us the
opportunity to establish the extent of crystalline silicate enrichment
in JFCs and the degree of JFC nucleus heterogeneity.

Future airborne instrumentation on the Stratospheric Observatory for
Infrared Astronomy (SOFIA) will cover mid-IR wavelengths in
spectroscopy mode \citep{becklin07}.  With the telescope executing
science operations at altitudes $\gtrapprox12.5$~km, SOFIA will be an
important resource for comet astronomy at 15--40~\micron, where
atmospheric absorption greatly interferes with observations at
ground-based observatories.  At these wavelengths, the spectral
resonances of crystalline silicates are more separated in wavelength,
and therefore more distinct, than the resonances near 10~\micron.
With 15--40~\micron{} spectra, we can accurately measure the
mineralogical makeup of silicate crystals in a comet's coma by
measuring the central positions and strengths of the crystalline
emission resonances.  In particular, the Fe to Mg ratio of silicate
crystals is better constrained when the longer wavelength peaks are
measured.  SOFIA will add vital observations to the few crystalline
measurements made at these wavelengths.

\section{Silicates in Jupiter-Family Comets}\label{sec:silicates}
\subsection{Properties of silicate grains and their thermal
  emission}\label{sec:silicate-properties}

Silicate minerals in comets have prominent emission features in the
mid-IR.  Molecular bonds between Si and O atoms have stretching and
bending modes that produce emission at 8--12~\micron{} and at
15--40~\micron, respectively.  The central wavelengths, relative
strengths, and shapes of these features are diagnostic of the mineral
composition.  \citet{wooden02} reviews the silicate minerals observed
in ground-based mid-IR spectra of comets:
[Mg$_y$,Fe$_{1-y}$]$_2$SiO$_4$ (olivine) and
[Mg$_x$,Fe$_{1-x}$]SiO$_3$ (pyroxene), where $x$ and $y$ range from 0
(Fe-pure: fayalite and ferrosilite) to 1 (Mg-pure: forsterite and
enstatite).  Cometary silicates are found in both amorphous
(non-stoichiometric) and crystalline structures.  An example of the
narrow thermal emission resonances attributed to silicate crystals,
and the broad emission attributed to amorphous grains is presented in
Fig.~\ref{fig:9p-decomp}.

Comet comae are optically thin to mid-IR radiation.  In order to
detect the thermal emission from silicate grains in a dust coma, the
grains must be in sufficient quantities to be detected above the
observational uncertainty of the smooth thermal continuum
\citep{hanner94}.  Aside from the amount, each grain's size,
composition, and structure play important roles in the strength of the
silicate emission, primarily by determining the grain equilibrium
temperature.  We discuss each of these properties (size, composition,
and structure) in turn.

First, dust grain size affects the grain temperature.  Sub-\micron{}
sized grains have higher temperatures than (solid) grains larger than
1~\micron{} of the same composition.  The temperature difference
arises from the dependence of grain emission efficiency on grain size.
All silicate dust grains greater than 0.1~\micron{} in size are
efficient absorbers of sunlight at visible wavelengths (the peak of
the solar output), but the sub-\micron{} sized grains are inefficient
emitters in the mid-IR.  Since grain radiative cooling at
$r_h\approx0.5-4$~AU is dominated by mid-IR radiation, the
temperatures of sub-\micron{} sized grains are warmer, which
subsequently increases their total thermal emission flux.

In contrast with grain size, which primarily affects the emission
efficiencies of comet grains, composition affects grain absorption
efficiencies.  Increasing the amount of Fe in a silicate grain
(decreasing Mg) increases the grain's absorption efficiencies at
visible and near-IR wavelengths, which in turn increases the grain's
equilibrium temperature \citep{dorschner95}.  The higher temperatures
cause Fe-rich grains to be easier to detect by spectroscopic means,
yet \citet{harker99} found the grain temperatures of amorphous
silicates in comet Hale-Bopp were best-modeled with roughly equal
amounts of Mg and Fe ($x = y = 0.5$).  Furthermore, \citet{harker99}
found the silicate crystals to be distinctly Mg-rich ($x = y =
0.9-1.0$).  These grain parameters have been commonly used by other
investigations \citep{moreno03, honda04, min05, harker07}.  However,
\citet{lisse06} found significant amounts of Fe-rich pyroxene in the
\di{} ejecta from comet 9P/Tempel (see \S\ref{sec:tempel} and
Fig.~\ref{fig:bar}).  In Fig.~\ref{fig:silicates-v-wave}, we summarize
the variation of the peak wavelength position with Mg content for
strong crystalline olivine and ortho-pyroxene peaks observed near
10~\micron{} from the analyses of \citet{koike03} and
\citet{chihara01}.

Finally, grain structure (i.e., porosity and shape) influences the
thermal emission from dust.  For two otherwise similar grains,
increased porosity will produce a warmer equilibrium temperature and,
in larger grains, a stronger 10~\micron{} silicate emission feature
\citep{hage90, xing97, min06, kolokolova07, voshchinnikov08}.  The
effect is due to insulating vacuum inside porous grains, causing
sub-domains to be increasingly thermally isolated from each other.  If
the porosity of an aggregate grain is more than 63\% then the
temperature begins to approach that of the monomer
($\sim$0.1~\micron{} sub-grains): at 80\% porosity the temperature of
the aggregate is only 4--6\% lower than a monomer \citep{xing97}.
Also, the shape of a porous aggregate determines whether larger porous
grains lose or retain their silicate resonant features
\citep{kolokolova07}.  The net result of increased porosity is a
thermal emission spectrum with a hotter temperature and a stronger
silicate feature compared to a similar sized compact grain.  Thermal
emission models for porous aggregates that include crystals
\citep[computed with the Discrete Dipole Approximation;][]{moreno03}
yield similar crystalline fractions as the less time-intensive
approach of modeling a grain distribution composed of discrete
crystals and mono-mineralogic porous grains \citep{harker02,
  harker04a}.

The above three grain parameters (size, composition, and structure)
all affect the observed \textit{strength} of a coma's silicate
emission feature, but each parameter also affects the \textit{shape}
of the feature.  When we consider the silicate emission feature's
shape, and the shape of the entire spectrum, we find that there are no
one-to-one degeneracies among the dust grain parameters, although
there are consistent trends.  For example, the silicate feature
strength can be raised by increasing the porosity of grains above
$\sim80$\% vacuum, which yields silicate features and temperatures
similar to their monomers \citep{xing97}.  Increasing the porosity of
dust grains in a coma also shifts the spectral slope of the
Rayleigh-Jeans side of the observed spectrum towards the blue (i.e.,
there are fewer, cool grains contributing to the flux at long
wavelengths).  A small portion of this change in spectral slope can be
accounted for by flattening the size distribution to increase the
relative number of larger, cooler grains.  The entire effect, however,
cannot be accounted for in this manner---this is evident when we
consider that the largest, highly porous grains have higher
equilibrium temperatures than the largest grains with more compact
shapes.  Oort cloud comet Hale-Bopp is an example of a comet that
displays an extremely strong silicate feature, and that fitting its
spectrum requires a size distribution that possesses both highly
porous grains and very small grains \citep[peak grain radius of
  0.15--0.2~\micron;][]{harker02, moreno03, min05}.  \citet{wooden02}
discusses how comet Hale-Bopp's grain porosity and grain size
distribution slope are each constrained by spectra.  In another
example, Oort Cloud comet C/2001 Q4 (NEAT) has a silicate feature
identical in shape to, but weaker in strength than, that of Hale-Bopp
\citep{wooden04}.  In fact, Q4's silicate feature strength varies on
short times scales and is best explained by a variable
silicate-to-amorphous carbon mass ratio \citep{wooden04, harker04b}.
Specifically, decreasing the silicate feature strength by a factor of
1.5 by removing the grains smaller than 0.6~\micron{} broadens the
shape of the silicate feature more than what is observed for Q4.
Therefore grain size effects cannot explain the observed variations in
the spectra of comet Q4.

With the above arguments in mind, we can consider the positive
detection of silicate features in many JFCs (Table~\ref{table:obs})
and draw some conclusions.  The weak silicate band emission in JFCs,
as compared to some OC comets, could be due to: 1) silicates may be
less abundant with respect to other dust components (e.g., amorphous
carbon, FeS) in JFCs than in OC comets; 2) silicate grains may be
larger in JFCs than in OC comets; 3) JFC dust may be less porous than
OC comet dust; or 4) a combination of 1, 2, and/or 3.  Out of the few
JFCs that have been studied in-detail, there is no clear agreement in
the silicate dust content of their comae.  \citet{kelley06apj} found
Jupiter-family comet 2P/Encke had a peak grain size distribution of
0.4--0.5~\micron{} and a sub-\micron{} silicate dust mass fraction of
$<31$\% ($3\sigma$ upper-limit).  \citet{hanner96} reasonably
reproduced the silicate feature of 19P/Borrelly to within 10\% using a
50-50 mixture of silicates and carbon.  Models of pre-\di{} spectra of
9P/Tempel favored a silicate dust distribution with a peak grain size
of 0.7--0.8~\micron{} for solid to somewhat porous grains
\citep{harker07}.  It may be that these three comets are truly
different in their compositions and grain size distributions, however
the models could be refined with new or improved constraints.  For
example, the comet Encke model fits could benefit from higher
sensitivity spectra at $\lambda<14$~\micron{} that better constrain
the grain temperatures and silicate compositions, and all three comet
models could be improved with constraints derived from scattered light
observations (\S\ref{sec:scattered}).  Thermal emission modeling has
the potential to constrain grain parameters, but only within the
limitations of a given data set.

A few OC comet thermal emission spectra have also been modeled.
Models of comets C/2001 Q4 (NEAT), C/2002 V2 (NEAT), and Hale-Bopp
reveal comae with smaller peak grain sizes (0.1--0.5~\micron) that are
dominated by crystalline silicate grains \citep[see][and references
  therein]{ootsubo07a}.  \citet{hanner94} find that weak silicate
emission in several OC comets is correlated with a weak dust
scattering continuum and weak 3~\micron{} thermal emission, altogether
suggesting a lack of small grains.  From the thermal emission studies
it is not clear which of the possible causes (1, 2, or 3) manifest the
differences between comet mid-IR spectra, but, from the limited
results above, OC comets seem to have an abundance of the smallest
grain sizes, when compared with JFCs.  In the next section we present
recent developments in models of light scattered by comet dust that
suggest grain porosity is the dominant property.

\subsection{Grain Porosity, Thermal Emission, and Light Scattered by
  Comet Dust}\label{sec:scattered}

Models and observations of light scattered by comet dust, in tandem
with mid-IR observations, are providing intriguing evidence that
supports grain structure as the dominant cause for the difference
between JFCs and OC comets.  In light scattering models, grain
structure refers to one of three types: solid grains; compact porous
aggregates, described by ballistic particle-cluster aggregation
(BPCA); or, fluffy (highly porous) aggregates, described by ballistic
cluster-cluster aggregation (BCCA) \citep{mukai92}.  Grain structure
affects the strength of the silicate feature, polarization, and
surface brightness radial profiles \citep[the latter due to the
  changing efficiencies of gas drag and the solar radiation
  force;][]{meakin88, mukai92, nakamura94}.

\citet{kolokolova07} delineate two types of dust comae: 1) comae
dominated by compact grains that are either solid, or compact porous;
or 2) comae dominated by fluffy aggregates.  In scattered light models
\citep{kimura03, kimura04, kimura06}, both grain types reasonably
reproduce the observed optical properties of comet dust \citep[for a
  review, see][]{kolokolova04}, but one observable breaks the
degeneracy: the relationship between the 10~\micron{} silicate
emission and the structure of the aggregates.

\citet{kolokolova07} show that the surface brightness distributions of
the dust and gas in comet comae vary from comet-to-comet.  They
suggest that gas-rich comets have dust comae that are more spatially
compact than their gas comae \citep[e.g., comet Encke,][]{jockers05},
and dusty comets have dust and gas comae with similar surface
brightness distributions.  Gas-rich comets have lower optical
polarization maxima than dusty comets, and the low optical
polarization is correlated with weak or absent 10~\micron{} silicate
features \citep{aclr96}.  Exploiting the correlation between low
optical polarization and dust surface brightness distribution, and
using recent observations that have shown the low optical polarization
is due to gas contamination in broadband filters \citep{kiselev04,
  jewitt04, jockers05}, \citet{kolokolova07} develop a physical
explanation to describe the correlation between dust-to-gas ratios and
the 10~\micron{} silicate emission feature.  Their argument is as
follows: 1) fluffy, porous aggregates have stronger or equivalent
10~\micron{} silicate emission features compared to compact, or solid,
aggregates (summarized above); 2) fluffy aggregates should be more
easily accelerated to the gas outflow velocity than compact grains,
primarily due to their larger surface areas per mass \citep{meakin88,
  nakamura94}; 3) in constrast, compact grains are not accelerated to
the gas expansion velocity, and dust comae consisting of compact
grains will be more concentrated around the comet nucleus than the
gas.  Altogether, the scattered light and the thermal emission
properties of comet dust suggest that grain porosity could be the
predominant characteristic that determines the strength of the comet's
silicate features.

\subsection{The influence of the nucleus on observed silicate band
  strength}\label{sec:nucleus} 

When assessing the dust content of comet comae, it is important to
remove the contribution of the nucleus from the total thermal emission
flux.  The nucleus can be a dominant contributor to the mid-IR flux of
JFCs, which are typically less active than their Oort Cloud
counterparts \citep{ahearn95}.  A large nucleus flux skews the
silicate band strength towards lower values.  To illustrate this fact,
we estimate the fraction of the mid-IR flux attributable to the
nucleus in a typical JFC observed at $r_h=1.8$~AU.  We will base our
model comet on surveys of JFC dust production rates \citep{ahearn95}
and nucleus sizes \citep{lamy04}.  We will also ``observe'' our model
comet at $r_h=1.0$~AU by appropriately increasing the dust production
rate and phase angle to show how the nucleus fraction varies with
heliocentric distance.  We then examine Table~\ref{table:obs} to
determine if the weak silicate features are actually a product of
large nuclei with low dust production rates.

Since visible light observations of comets are more readily available
than mid-IR observations, it is advantageous to use the visible
observations to estimate the coma brightness of a typical JFC.  The
visible light proxy for dust production is the quantity \afrho, where
$A$ is the Bond albedo of the dust, $f$ is the dust areal filling
factor in the field-of-view, and $\rho$ is the projected radius of the
field-of-view, typically expressed in units of cm \citep{ahearn84}.
The filling factor may be expressed as $f = N\sigma \pi^{-1}
\rho^{-2}$, where $N\sigma$ is the total cross-section of the grains
in cm$^2$.  Technically, $f$ depends on the extinction cross-sections
of the grains and not on the geometric cross-sections
\citep{ahearn84}.  For comae with constant isotropic outflows,
\afrho{} is independent of aperture size.

The thermal emission, $F_\lambda$ in units of \wcmum, from a
collection of isotropically emitting isothermal grains is
\begin{equation}
  F_\lambda = (1 - \bar{A})~\pi B_\lambda(T)~f \frac{\rho^2}{\Delta^2},
\label{eq:grain1}
\end{equation}
where $\bar{A}$ is the mean bolometric albedo of the dust
\citep[$\approx0.32$;][]{gehrz92}, $B_\lambda(T)$ is the Planck
function in units of \wcmumsr{} evaluated for the grain temperature
$T$ and at a wavelength of $\lambda$, and $\Delta$ is the
comet-observer distance in cm.  It is useful to express
Eq.~\ref{eq:grain1} in terms of $Af\rho$ and $\theta$, where $\theta =
206265~\rho~\Delta^{-1}$ is the angular size of the aperture in units
of arcseconds,
\begin{equation}
  F_\lambda = \frac{(1 - \bar{A})}{A(\alpha)}~\pi
              B_\lambda(T)~\frac{\theta}{206265}~
              \frac{A(\alpha)f\rho}{\Delta}.
\label{eq:grain2}
\end{equation}
Here, we explicitly write the phase angle dependence of the albedo as
$A(\alpha)$; the thermal emission term, $1-\bar{A}$, is independent of
phase angle.  Visual albedo measurements of Jupiter-family comets are
rare.  We found that only one comet, 21P/Giacobini-Zinner, has been
studied, and it has a low albedo: $A(64\degr)=0.07-0.15$
\citep{telesco86}, $A(22\degr)=0.11$ \citep{pittichova08}.  In the
near-IR, JFCs have albedos equivalent to or lower than Oort Cloud
comets \citep{campins82, hanner89}.  Since comets are typically
observed at low-to-moderate phase angles, we have chosen
$A(\alpha<60\degr)$ to be 0.15, which is slightly under the apparent
mean $A(\alpha)$ of several Oort Cloud comets \citep[see Fig.~1
  of][]{kolokolova07}.  The mid-IR coma flux is fairly sensitive to
$A(\alpha)$, therefore measured values should be used when possible.
Assuming a value of 0.15 introduces a factor of 2 uncertainty in the
mid-IR flux when we consider the observed range is $\approx0.1-0.25$
\citep[excluding comet Hale-Bopp;][]{kolokolova04}.  We take the
ensemble average dust temperature to be 10\% warmer than the blackbody
temperature, $T_{BB}$, at the observed $r_h$,
\begin{equation}
  T\approx 1.10~T_{BB}= \frac{306 \mbox{\textrm{K}}}{\sqrt{r_h}},
\end{equation}
which is within the typical range (0--30\%) observed in many comets
\citep{gehrz92, hanner96}.

The near-Earth asteroid thermal model \citep{harris98} has proven to
be sufficient for modeling the thermal emission from comet nuclei,
although with some modifications.  \citet{groussin04} suggests an
IR-beaming parameter of 0.85 for comets, based on the surface models
of \citet{lagerros98}.  In the formalism of the near-Earth asteroid
thermal model, the beaming parameter contributes to the observed
effective temperature of the nucleus.  The value of 0.756 has been
favored for large, main-belt asteroids \citep{lebofsky86}, but
equivalent or larger values, up to $\approx1.0$, have been
experimentally determined with infrared observations of comet nuclei
\citep{fernandez00, kelley06apj, fernandez06, groussin07}.  We adopt
0.85 throughout our paper.  We also use a geometric albedo of 0.04, a
typical value for comet nuclei \citep{lamy04}.

\citet{ahearn95} surveyed the optical properties of 85 comets, of
which 34 belong to the Jupiter-family class.  Their median observed
\afrho{} value for JFCs is 110~cm at a median $r_h$ of 1.8~AU.  We
derive a median nucleus radius of 1.8~km for the 34 JFCs using the
radii listed by \citet{lamy04} and updated values for 81P/Wild and
9P/Tempel from \citet{brownlee04} and \citet{ahearn05}.
Table~\ref{table:estimate} lists the nucleus-to-coma flux ratios at
10~\micron{} for various observing parameters of our median JFC
($\afrho=110$~cm, $r_h=1.8$~AU, $R_n=1.8$~km).  Of course, our
fictional ``median JFC'' parameters are not representative of any
particular JFC, and that spectra of each comet should be considered on
a individual basis.  Table~\ref{table:estimate} also lists estimates
for the median JFC observed at $r_h=1.0$~AU by scaling the \afrho{}
value up by a factor of $1.8^{2.3}$, as suggested by \citet{ahearn95}.

A bright nucleus may affect derived coma parameters, the most
immediate of which is the silicate band strength.  It is clear from
Table~\ref{table:estimate} that the nucleus flux can comprise a
significant amount of the total flux measured in mid-IR spectra, as
high as 80--90\%, especially when observing comets with low dust
production rates ($\afrho\approxeq100-200$~cm) using the narrow-slits
(0.4--1.0\arcsec{} wide) on ground-based instrumentation.  In this
case, if the silicate band strength appears to be between 1.10 and
1.30, then the nucleus subtracted (corrected) silicate band strength
would lie between 1.50 and 3.0!  With a larger slit, the nucleus could
be 10--30\% of the flux.  The corrected silicate band strength would
then increase to 1.11--1.43.  The latter small changes in the silicate
band emission of a JFC may not be significant when comparing a JFC to
comet Hale-Bopp, which had a silicate emission feature that ranged
from 3.0 to 4.0, but are important when assessing the properties of
JFCs as a whole.

Removing the nucleus is relevant to grain thermal emission modeling,
even in comets with no discernible silicate emission feature.  Comet
nuclei have temperature distributions unlike collections of isothermal
grains (the dust coma).  Comet nuclei have low thermal inertia
\citep[e.g.,][]{groussin07}, therefore, the sub-solar point on a
slowly rotating nucleus will be hottest, and the temperature decreases
as the sun-zenith angle ($\theta_\sun$) increases (in the near-Earth
asteroid model, $T\propto\cos^{1/4}{\theta_\sun}$).  In contrast, the
thermal propagation timescales in small grains are fast and they
quickly equilibrate to a single temperature---a temperature that is
cooler than the effective temperatures of comet nuclei.  Thus, the
shortest wavelengths of a thermal emission spectrum will include the
highest fraction of flux from the nucleus \citep[e.g., see the
  reduction of the 2P/Encke spectra in][]{kelley06apj}.  If the
nucleus emission is not removed, a coma modeler would find an
increased number of the smallest grains (the grain sizes that are
constrained by the shortest wavelengths) over a best-fit model that
excludes the nucleus flux.

We now refer to Table~\ref{table:obs} and discuss our estimates of how
the thermal emission from each comet's nucleus affects the observed
silicate band strength.  \citet{hanner96} considers the size of the
comet nucleus when computing the silicate band strength of comet
4P/Faye, so no correction is necessary.  Based on images of the
nucleus of comet 19P/Borrelly \citep{buratti04}, the $8.0\times3.2$~km
prolate nucleus contributes 14--36\% of the flux in the spectrum of
\citet{hanner96}.  The silicate emission feature strength is reported
by \citet{hanner96} to be $\approx1.25$.  Their plot of the data,
however, suggests that $\approx1.11$ is a more appropriate value.
Removing the nucleus flux increases the strength to 1.13--1.19.
\citet{stansberry04} estimate the silicate band strength of comet
29P/Schwassmann-Wachmann to be approximately 1.10.  We use their model
nucleus parameters and derive a corrected band strength of 1.19.
\citet{sitko04} observed 69P/Taylor and find the silicate band
strength to be 1.23.  The radius of 69P is estimated to be
$2.1\pm0.6$~km by \citet{tancredi06}, which corresponds to 4--19\% of
the thermal emission in the \citet{sitko04} mid-IR spectrum.  The
result is a minor increase in the the silicate band strength to
1.24--1.28.  The silicate band strength of comet 103P/Hartley (1.20)
measured by \citet{crovisier00} requires no correction due to the
large aperture of the ISOPHOT instrument ($24\arcsec\times24\arcsec$),
and the small nucleus size, $R_n=0.71\pm0.13$~km \citep{groussin04}.
We derive a nucleus contribution of $\leq1$\% for comet 103P.
Finally, the observation of 103P/Hartley by \citet{lynch98} requires a
minor correction, resulting in an increase of the measured silicate
band strength from 1.15 to 1.22, which matches the ISOPHOT derived
silicate band strength.

\section{Comet 9P/Tempel and \di}\label{sec:tempel}

Comet 9P was the primary target of the \di{} mission---a mission
designed to impact a comet nucleus to determine its underlying
strength, structure, and composition.  In all ways, comet 9P/Tempel is
considered to be a typical JFC \citep{ahearn07}.  The \di{} spacecraft
delivered a 370~kg impactor into the nucleus of 9P/Tempel at a
relative velocity of 10.3~\kms.  The fly-by spacecraft observed the
event from a distance of 500~km \citep{ahearn05}.  The impact
excavated $\sim10^5-10^6$~kg of ice and dust from the surface and
interior of the comet \citep{sugita05, harker05, kueppers05,
  keller07}.

Some of the most detailed mid-IR spectra of any JFC were obtained of
comet 9P/Tempel in support of the \di{} mission.  Three investigations
have been presented: 1) \spitzer{} observations of the pre- and
post-impact coma (approximately 30 and 60~min after impact) at
5--40~\micron{} \citep{lisse06}; 2) Gemini-N observations of the pre-
and post-impact coma (approximately 7~min cadence for 3~hr, and
additional spectra 24~hr after impact), primarily at 8--13~\micron{}
\citep{harker07}; and 3) a Subaru observation of the post-impact coma
(3.5~hr after impact) at 8--13~\micron{} \citep{ootsubo07b}.

\subsection{The Pre-\di{} Dust Coma}

The \di{} spacecraft studied the ambient coma and nucleus of comet 9P
and provided excellent context for observations from other spacecraft
and ground-based observatories.  \citet{farnham07} analyzed images of
9P/Tempel taken by the \di{} spacecraft and found that the pre-impact
dust coma has three prominent jets, the strongest of which originates
near comet 9P's south pole.  \citet{feaga07} observed \coo{} and
\water{} gas in the coma and find that \coo{} is preferentially
ejected from the southern pole (correlated with the dust) and that
\water{} is preferentially ejected in the sunward direction (on the
northern hemisphere at the time of observation).  The \coo{} and
\water{} asymmetries gives strong evidence for either differentiation
of ices in the nucleus, or primordial heterogeneity.  Comet jets will
typically be unresolved in mid-IR observations, but it would be
interesting to investigate whether dust from 9P's discrete sources
propagate into separate coherent coma or tail structures
\citep{farnham07}, and if the dust properties of these structures
could be assessed using the high signal-to-noise spectral map of comet
9P taken with the \spitzer/IRS instrument.  For the purposes of this
review, we will assume analyses of mid-IR observations of the
pre-\di{} coma probe a global average of the dust properties of this
comet.

\citet{lisse06} present a pre-impact \spitzer{} spectrum of 9P/Tempel,
but a detailed analysis of this spectrum is not given.  In
Fig.~\ref{fig:9p-spitzer} we present the pre-impact spectrum taken by
\spitzer{} on 03 July 2005.  A model nucleus has been removed, and a
silicate feature (strength of $\approx20$\%) is clearly observed at
8--12~\micron.  We estimate the silicate band strength to be in the
range 1.19--1.25.

\citet{harker07} describe their pre-impact Gemini spectra as
``essentially smooth and nearly featureless'' with a silicate band
strength close to 1.0 (no silicate emission), although an uncertainty
is not presented (the nucleus also amounts to 80\% of their pre-impact
flux).  We estimate the silicate band strength to be $1.2\pm0.2$
(after nucleus subtraction).  A spectrum taken in May 2005, about 1.5
months prior to the \di{} encounter, did show clear silicate emission
(we estimate the strength to be $1.50\pm0.18$).  In May 2005 when dust
from the southern jet dominates the coma \citep{farnham07}, the dust
mineralogy is dominated by solid (zero porosity) amorphous pyroxene
silicates with a peak grain size of 0.7~\micron, whereas in July 2005
pre-impact the ambient coma is comprised of moderately porous (46\%
vacuum for a 10~\micron{} grain) amorphous carbon and amorphous
olivine silicates with a slightly larger peak grain size of
0.8~\micron.  Note that for a 10~\micron{} radius grain, the silicate
feature strength increases from negligible to $\approx1.2$ as the
porosity changes from solid (0\% vacuum) to moderately porous
($\approx$50\% vacuum) \citep{harker02, harker07}.  In summary, JFC
comae with low dust production rates or measured through small slits
may at first appear to have ``featureless'' IR spectra (especially at
low signal-to-noise ratios), but modeling after nucleus removal can
reveal micron-sized solid or 10~\micron{} porous silicate grains, as
is the case for pre-impact comet 9P \citep[Figs.~2 and 5
  of][]{harker07}.

\subsection{The Post-\di{} Dust Coma}

Post-impact mid-IR spectra were dramatically different from the
pre-impact spectra (Fig.~\ref{fig:9p-spitzer}).  Mid-IR spectra of the
ejecta show the strong resonances of sub-\micron{} sized crystalline
silicates of both olivine and pyroxene grains \citep{lisse06,
  harker07, ootsubo07b}.  This discovery is perhaps the most immediate
result of the mid-IR investigations.  A \spitzer/IRS observation of
the 10~\micron{} silicate feature (0.64~hr after impact) is presented
in Fig.~\ref{fig:silicates-v-wave}.  We find the post-impact silicate
band emission strength to be near 2.0 (both the Gemini and \spitzer{}
observations; see column 6 in Table~\ref{table:obs}), the strongest
feature ever observed in a spectrum of a JFC.  The derived
crystalline-to-amorphous silicate sub-\micron{} mass ratios range from
approximately 0.2 to 4 \citep{harker05, harker07, lisse06,
  ootsubo07b}.  The values depend on slit width, orientation, and time
of observation, which could easily explain the wide range of values.
The pre-impact crystalline-to-amorphous silicate ratio has not been
significantly constrained \citep[$\lesssim1.6$, $1\sigma$
  upper-limit][]{harker07}.

Only in two or three other JFCs, 78P/Gehrels \citep{watanabe05},
73P/Schwass\-mann-Wach\-mann \citep{harker06b}, and, possibly,
103P/Hartley \citep{crovisier00}, have detections of crystalline
silicates been reported.  We may also include the Centaur-like JFC
29P/Schwassmann-Wachmann \citep{stansberry04}.  For comparison with
9P, \citet{watanabe05} found a crystalline-to-amorphous silicate ratio
of $\approx0.9$ for 78P/Gehrels.  The crystalline-to-amorphous
silicate ratios of 78P and the \di{} ejecta are significantly greater
than that found in the interstellar medium, $\leq 0.022$,
\citep{kemper04, kemper05}, and similar to or smaller than the
crystalline-to-amorphous ratios measured in the comae of some Oort
Cloud comets \citep[2--3 for comets Hale-Bopp, C/2001 Q4 (NEAT), and
  C/2002 V1 (NEAT);][]{harker02, harker04a, wooden04, ootsubo07a}.

Aside from the crystals, \citet{harker07} find that the \di{} ejecta
was rich in amorphous pyroxene, in contrast with the amorphous olivine
and amorphous carbon dominated pre-impact coma in July 2005, but
similar to the higher dust production rate coma in May 2005.  In the
impact-induced coma, crystalline olivine is correlated with the
amorphous pyroxene dust, suggesting these minerals originated from the
same reservoir (the \di{} crater).  \citet{harker07} also found a
population of amorphous carbon grains traveling at a projected speed
of 700~m~s\inv{}, equivalent to the highest speed ejecta seen by other
observers \citep[see, e.g.,][]{keller07}.  \citet{schultz07} interpret
high speed ejecta as originating from the upper 5~m of the nucleus,
whereas slower ejecta is from deeper layers.  \citet{harker07} and
\citet{kadono07} conclude that the surface of the nucleus is rich in
sub-\micron{} amorphous carbon grains.

With the broad wavelength coverage of \spitzer{} (5--40~\micron),
\citet{lisse06} were able to identify a number of minerals that have
not been previously reported in cometary spectra, namely Fe-rich
silicates, metal sulfides, phyllosilicates, and carbonates.  To
emphasize the benefit of the long wavelength coverage, we assembled a
bar chart consisting of 9P/Tempel's dust components
(Fig.~\ref{fig:bar}).  The figure highlights those minerals that in
principle are difficult to constrain without the 5--40~\micron{}
\spitzer{} spectra---these components add up to one-third of the total
dust surface area.  \citet{lisse06} find that 8\% of all the silicates
by emitting surface area are in the form of the Fe-rich hydrated
mineral nontronite, and 5\% of the total surface area is in the
Mg-rich carbonate magnesite, or the Fe-rich carbonate siderite.  In
\S\ref{sec:hydrated}, we discuss implications of the presence of
hydrated minerals on our knowledge of comet nuclei.

\citet{harker07} and \citet{ootsubo07b} assumed a Mg-rich composition
for their crystalline silicate grains, which allowed adequate fits to
most of their spectra ($\chi^2 \lessapprox 1.0$).  \citet{lisse06}
simultaneously fit both Mg-rich and Fe-rich silicate minerals to their
spectra and find significant amounts of Fe-rich silicate crystals.  By
number, the olivine Fe-/Mg-rich ratio is 0.3, and the pyroxene
Fe-/Mg-rich ratio is 3.1.  The difference between the Fe-/Mg-rich
ratio in these two major silicate groups is intriguing.  Both comets
9P and Hale-Bopp have high crystalline silicate fractions, but their
specific minerals are different.  Only the Mg-rich species were
present in comet Hale-Bopp \citep{wooden99, bradley99, harker02}.
Furthermore, an 11.25~\micron{} peak in the spectra of other OC comets
indicates Mg-rich grains \citep{hanner94}.  As discussed below
(\S\ref{sec:crystals}), if aqueous water was not present in the comet
nucleus (\S\ref{sec:hydrated}), then the Fe content of silicate
minerals may be a gauge of the amount of water vapor present in the
solar nebula during silicate crystal formation.  Future observations
and analysis of crystals in Jupiter-family and other comets from
\spitzer{} and SOFIA are vital to our understanding of the origin and
evolution of dust in our solar system.

\section{The Implications of Hydrated Minerals and Fe-Rich Crystals on
  Dust Formation and Comet Evolution}\label{sec:9p-discussion}
\subsection{Hydrated Minerals}\label{sec:hydrated}
The suggestion that Fe-rich crystalline silicates, phyllosilicates,
and carbonates are components of comet dust has consequences on our
understanding of dust formation and cometary evolution.  Significant
amounts of liquid water would be required if these minerals were to
have formed in 9P/Tempel's nucleus \citep{lisse06}.  This would upset
the common understanding that the interiors of comets remained at low
temperatures after formation \citep{meech04}.  Carbonates are thought
to form by aqueous alteration inside asteroids \citep{armstrong82,
  zolensky97}, because under steady-state conditions in the PPD
mid-plane, the nebular concentration of CO$_2$ is too low
\citep{lewis79}.  The (anhydrous) mineral olivine could have been
aqueously altered to become the phyllosilicate serpentine in
water-rich shocks in the solar nebula phase \citet{ciesla03,
  fegley89}.  In contrast to the relatively simple incorporation of
the OH$-$ radical into olivine to form serpentine, the formation of
smectite requires mobilization of refractory element cations including
Ca, Mg, and Si.  The gas transport of refractory elements requires
very high temperatures and this seems inconsistent with a shock-wave
producing low temperature mineral phases such as phyllosilicates
(A. Krot, private communication).  Hence, phyllosilicates other than
serpentine probably formed in the interiors of asteroids where
pressures were higher than in the nebula \citep{alexander89,
  brearley06, scott05}.

To aqueously alter minerals in the comet nucleus, some mechanism must
warm the comet interior enough to melt or sublime water ice.  Deeper
than a few km, heating by $^{26}$Al is sufficient to crystallize
amorphous water ice, but much of the excess heat is quickly dissipated
due the the higher heat conductivity of crystalline ice (about 20
times the heat conductivity of amorphous ice).  Only in comets with
radii greater than 10~km can the temperature continue to rise to allow
for liquid water to form \citep{prialnik95, prialnik99, prialnik08,
  merk06, mckinnon08}.  The interior may be high enough to form
phyllosilicates by aqueous alteration of silicate mineral grains in
comet dust, but not hot enough to dry out the phyllosilicates to form
FeO-rich crystalline silicates, as has been suggested to occur in
asteroids \citep{krot95, krot00}.  In addition to phyllosilicates,
other minerals that characterize parent-body aqueous alteration
include hydroxides, hydrated sulfides, sulfates, oxides, and
carbonates \citep{zolensky97}.  These secondary minerals are notably
absent from comet 1P/Halley \citep{jessberger88, jessberger99} and
\stardust{} \citep{brownlee06, zolensky06, zolensky07}.  The
\citet{lisse06} analysis of comet 9P suggests low concentrations of
carbonates and the phyllosilicate nontronite (a smectite), at the
$\approx$5--10\% level in comet 9P, but the identification of these
minerals in spectra of other comets is still under study
\citep{woodward07, dbm07, lisse07}.  In an analysis of spectra of
comet Hale-Bopp, \citep{wooden99} set a tight constraint of
$\leq 1$\% on the phyllosilicate montmorilonite (a smectite).  Note
that radiogenic temperatures in comet nuclei are not sufficiently
high, nor the conditions ``dry'' enough, to anneal and crystallize
amorphous refractory silicate minerals in comet dust.  As described
before, crystalline grains appear to have formed prior to the
accretion of comet nuclei.

\subsection{Fe-Rich Silicate Crystals}\label{sec:crystals}
Since the discovery of crystalline silicates in comets
\citep{campins89}, it has been recognized that some comets exhibit
mid-IR spectral features attributable to crystalline silicates,
typically crystalline olivine.  The identification of Mg-rich
crystalline olivine at 11.2~\micron{} is secured by the \iso{} SWS
spectrum of comet Hale-Bopp that shows the distinct longer wavelength
features \citep{crovisier00}.  On the other hand, some OC comets do
not exhibit crystalline silicate emission features, although how much
of this dichotomy is due to observational uncertainties will not be
known until an analysis of a large number of comets is presented.
Crystalline silicates are found in such high abundances in some comets
that the crystals cannot be relics from the interstellar medium, where
the crystalline fraction is $\leq 2.2$\% \citep{kemper04,
  kemper05}.  For example, the crystalline-to-amorphous sub-\micron{}
mass ratio in the coma of the OC comet Hale-Bopp was 1.5--3.7
\citep{harker02, harker04a}.  Recent observations have shown that JFCs
also have crystals in significant fractional abundances: $\sim$0.3 for
Deep Impact-induced inner coma (3\arcsec) of 9P/Tempel
\citep{harker05, harker07, sugita05} and $\sim$0.8 for the Deep-Impact
larger coma \citep[10\arcsec][]{lisse06}, $\sim$0.9 for 78P/Gehrels
\citep{watanabe05}, and qualitatively described as less than in
Hale-Bopp is 73P-B/Schwassmann-Wachmann and 73P-C/Schwassmann-Wachmann
\citep{harker06b, sitko06irs-iauc}.

Exactly what conditions led to the formation of crystalline silicates
and their transport to the comet forming zone is not known, but the
high Mg-content of crystalline silicates found in comets requires
significant processing of proto-solar nebula amorphous silicates
\citep[see a review by][]{wooden07}.  Based on a legacy of solar
nebula condensation models developed to explain minerals in chondrules
and chondrites, Mg-Fe amorphous silicates evaporated and re-condensed
at $\sim1450$~K as extremely Mg-rich silicate crystals and separate Fe
grains in what is thought to be a typical ``dry'' solar nebula
\citep{krot00, weinbruch00}.  In addition, annealing, the process of
heating an amorphous grain enough to crystallize it, may have been an
important process.  Based on a wide variety of experiments designed to
create amorphous silicates and anneal them into crystals at
$\sim$1000~K \citep[see a review by][]{wooden05}, Mg-rich amorphous
silicates anneal to Mg-rich crystalline silicates, Fe-rich amorphous
silicates anneal to Fe-rich crystalline silicates, and Mg-Fe amorphous
silicates anneal to Fe-bearing crystalline silicates
\citep{brownlee05} but the Fe-content in the crystals begins low (0.1)
and increases (to 0.2) as the crystallization completes
\citep{murata07}.  The annealing products and activation energies
somewhat depend on how the amorphous silicate starting materials are
made in the laboratory \citep[see the review by][]{wooden05,
  murata07}.  The major issue with annealing to create cometary
crystals is that Mg-rich amorphous silicates are not seen in cometary
interplanetary dust particles or chondritic materials, so there is no
reservoir with which to anneal to Mg-rich crystals.  Instead,
annealing of Mg-Fe amorphous grains to form Mg-rich crystals must
occur in a low oxygen fugacity gas (i.e., when water is depleted
relative to molecular hydrogen) where Fe can be reduced (removed by
interdiffusion) \citep{davoisne06, wooden07}.

The small number abundance of Fe-rich crystals in cometary materials
can be explained either by annealing without Fe reduction
\citep[e.g.,][]{brownlee05} or by late-time condensation from the gas
phase in a wet nebula \citep[H$_2$O/H$_2$ $\approx$100--1000 times
  enhanced over a `dry' nebula;][]{palme90}.  Interpretation of
\stardust{} samples of comet 81P/Wild, where Fe-rich crystalline
silicates are found in small numbers \citep{zolensky07} but where
secondary aqueous alteration products (e.g., layer lattice silicates)
are not found \citep{zolensky06}, the gas phase condensation of
moderately Fe-enriched to Fe-rich silicates is a more consistent
scenario than parent body aqueous alteration.  Moreover, we suggest
that the condensation of Fe-rich crystalline silicates from a hot,
water-rich nebula at times late enough for 90\% of the Mg to have
condensed into solids \citep{palme90} may explain the presence of
Fe-rich crystalline silicates in both comets and in some chondrites
that otherwise do not show extensive signs of parent body aqueous
alteration.

As discussed above, Fe-rich crystalline silicates can condense from
the gas phase when the oxygen fugacity is high.  In order to have both
Mg-rich and Fe-rich crystalline condensates, the water content must be
variable.  Water variability could come from inward migration of icy
planetesimals into the hot inner-zones of the PPD \citep{cuzzi04}.
Therefore, the differences between the comets with Mg-rich crystals
and comet 9P/Tempel with Fe-rich crystals may be a matter of the epoch
of each comet's formation, and the efficiency of radial transport in
and out of the comet forming zones.  Studying the abundance of Fe-rich
crystalline silicates in bodies that have not undergone internal
aqueous alteration (i.e., comets) may aid our understanding of the
transport of water from beyond the frost line into the inner solar
nebula, and the subsequent enrichment of the outer solar system with
condensates that formed in a water rich environment.

\section{Summary}\label{sec:summary}

We reviewed ground-based and space-based mid-IR spectra of
Jupiter-family comets taken over the past 25 years.  These
observations, including spectroscopy of 9P/Tempel during the \di{}
encounter and early \spitzer{} results, allow us to draw some
conclusions on the nature of dust in JFCs: that, 1) JFCs have weak
silicate emission features, roughly 20--25\% over the continuum,
similar to the lowest values observed in Oort Cloud comets; 2) the
weak silicate emission could be due to low silicate content relative
to other grain species (i.e., amorphous carbon and FeS), or due to the
preponderance of large, compact grains ($\gtrsim1$~\micron); 3) three
JFCs show evidence for crystalline silicates in their mid-IR spectra
(four if we include the Centaur-like 29P/Schwassmann-Wachmann)---the
existence of crystals, also found in OC comet spectra, indicate high
temperature processes altered some dust in the proto-planetary disk;
4) JFC nuclei can have heterogeneous dust properties, the evidence for
which is presented in studies of the \di{} encounter with comet
9P/Tempel; 5) the evidence for hydrated minerals in the coma of
9P/Tempel is perplexing when we consider that comet interiors (for
$R\leq10$~km) likely remained well below the melting point of water
ice; and, 6) the crystalline silicate forming zones of the solar
nebula may have had a variable water content (i.e., variable oxygen
fugacity) , which could lead to the apparent dichotomy of comets:
comets that predominantly consist of Mg-rich crystals (e.g., comet
Hale-Bopp), and comets with significant amounts of Fe-rich crystals
(comet 9P/Tempel).

We identify several opportunities with which we may improve our
understanding of dust in JFCs: 1) increase the number of observed JFC
mid-IR spectra; 2) continue to model the thermal emission spectra of
JFCs in order to ascertain the nature of their weak silicate features;
3) continue to compare models of light scattered by comet dust with
thermal emission models to significantly constrain the structure of
dust grains; and 4) employ the remarkable capabilities of 8--10~m
class telescopes and SOFIA to improve our understanding of the amount
and mineralogy of silicate crystals in comets, especially their Mg and
Fe contents.  We hope to answer: Why do JFCs and some OC comets have
weak silicate emission features?  How common are Fe-rich silicates and
hydrated minerals in JFCs?  Is the typical JFC heterogeneous in dust
composition, and does that heterogeneity arise from the primordial
aggregation of planetesimals?

\section*{Acknowledgments}
\begin{figure}
\centering
\includegraphics[width=0.8\textwidth]{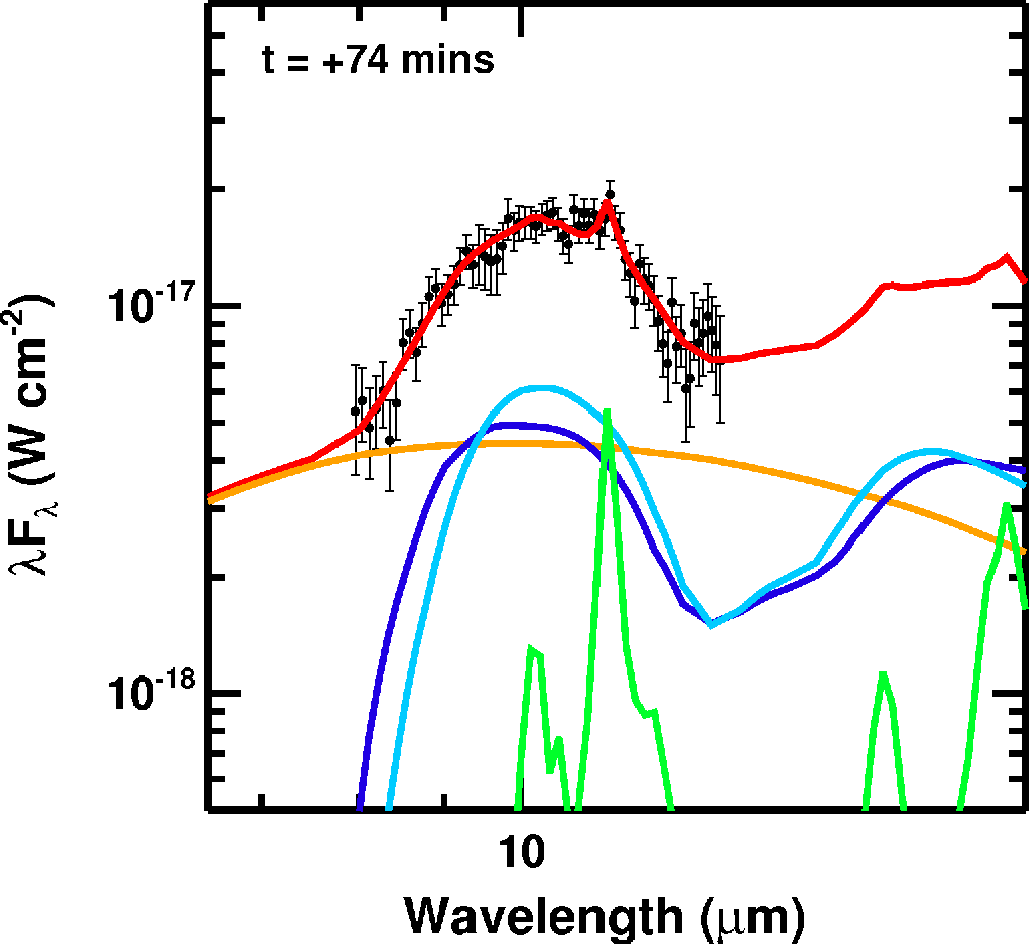}
\caption{Model spectra of the grain components detect in post-\di{}
  observations of 9P/Tempel from \citet{harker07}.  The
  Gemini-N/Michelle spectrum (\textit{black circles}) is centered on
  the nucleus and is taken 74~min after impact.  The thermal
  contribution of the nucleus is shown as an \textit{orange line}.
  Superposed on the spectra is the total model spectral energy
  distribution (\textit{red line}). The model is decomposed into its
  constituent dust components: amorphous pyroxene (\textit{blue
    line}), amorphous olivine (\textit{cyan line}), and crystalline
  olivine (\textit{green line}).  Notice the contrast between the
  broad spectral features of the amorphous silicates with the narrow
  resonances of crystalline olivine.
 \label{fig:9p-decomp}}
\end{figure}

\begin{figure}
\centering
\includegraphics[width=0.6\textwidth]{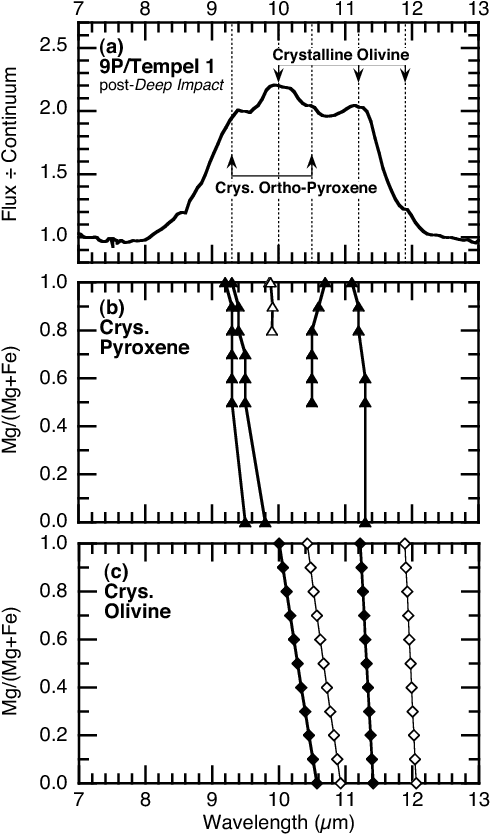}
\caption{The variation of selected resonance wavelength positions with
  Mg content in the 10~\micron{} region for crystalline olivine
  \citep{koike03} and crystalline ortho-pyroxene \citep{chihara01}.
  \textit{Top}| A post-\di{} \spitzer/IRS spectrum of comet 9P/Tempel
  (1~hr after impact) normalized by a scaled Planck function fit to
  the spectrum at 8 and 12~\micron{} (as described in
  \S\ref{sec:obs}).  The wavelength positions for silicate crystalline
  peaks observed in comets C/1995~O1 (Hale-Bopp) \citep{wooden99},
  C/2001~Q4 (NEAT) \citep{wooden04}, and 9P/Tempel \citep{harker07}
  spectra are marked with \textit{vertical dotted-lines}.
  \textit{Center and Bottom}| \textit{Filled symbols} mark the central
  wavelengths of strong silicate resonances as they vary with Mg
  content, \textit{open symbols} mark weaker silicate resonances.  The
  \textit{y-axis} values range from 0.0 (Mg-poor, Fe-rich) to 1.0
  (Mg-rich, Fe-poor) corresponding to the $x$ and $y$ parameters in
  each mineral's chemical composition: [Mg$_y$,Fe$_{1-y}$]$_2$SiO$_4$
  (olivine) and [Mg$_x$,Fe$_{1-x}$]SiO$_3$ (pyroxene).
 \label{fig:silicates-v-wave}}
\end{figure}

\begin{figure}
\includegraphics[width=\textwidth]{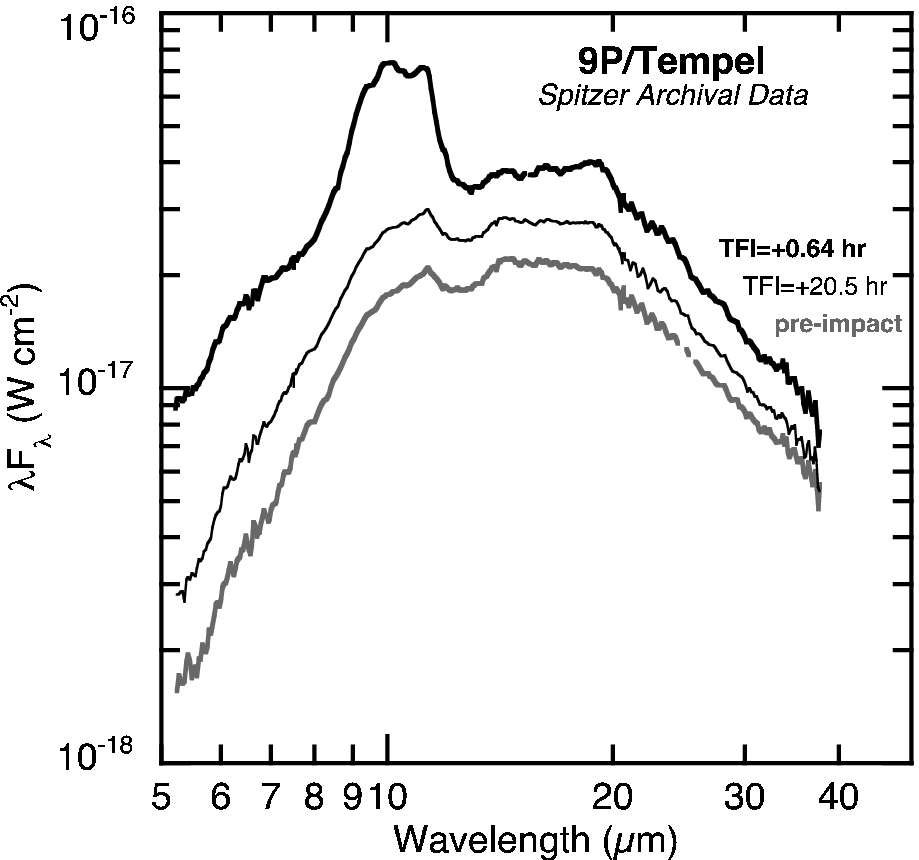}
\caption{Pre-impact spectrum of comet 9P/Tempel taken 03 July 2005 UT
  (\textit{gray line}), and two spectra taken 0.64 (\textit{thick
    black line}) and 20.5~hr (\textit{thin black line}) after impact.
  The contribution from a 3.3~km nucleus has been removed.  Notice the
  broad amorphous silicate emission at 8--12 and 15--20~\micron{} in
  the pre-impact spectra.  Also note the strong crystalline resonances
  at 9 and 11~\micron{} in the first post-impact
  spectrum.  \label{fig:9p-spitzer}}
\end{figure}

\begin{figure}
\includegraphics[width=\textwidth]{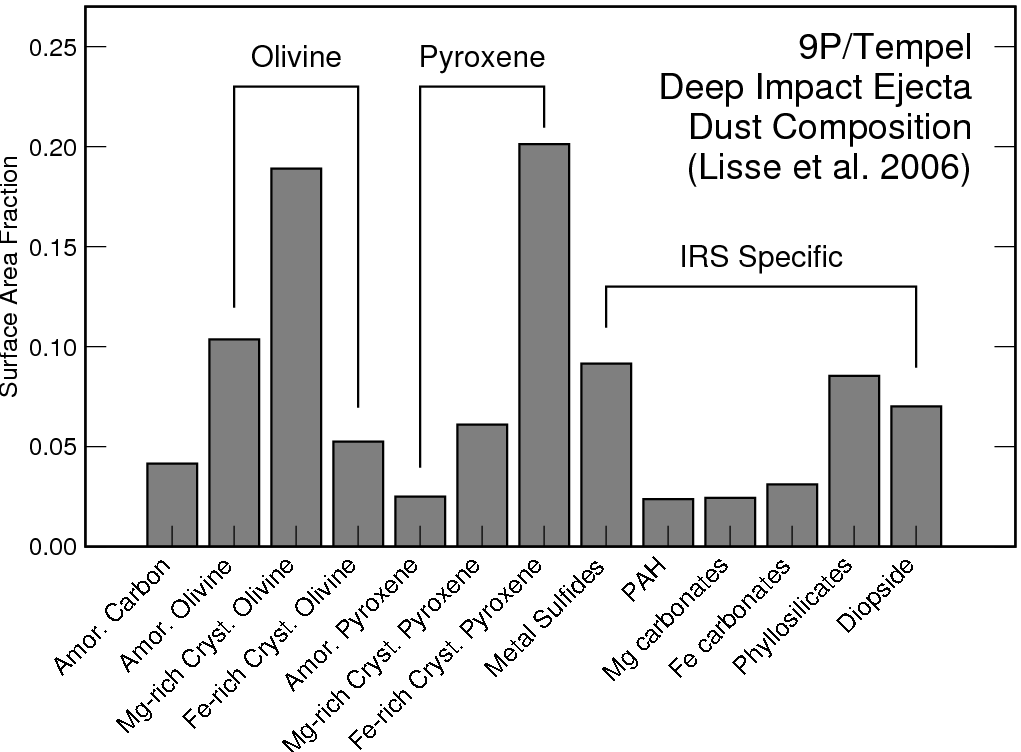}
\caption{Grain components identified in the mid-IR in post-\di{}
  \spitzer{} spectra of comet 9P/Tempel \citep{lisse06}.  In order to
  emphasize the importance of space-based infrared spectra, we have
  highlighted those components that have only been identified through
  reduced $\chi^2$ fits to the \spitzer/IRS 5--35~\micron{}
  observations. \label{fig:bar}}
\end{figure}

The authors gratefully acknowledge D.~E. Harker for providing
Fig.~\ref{fig:9p-decomp} and Gemini spectra of comet 9P/Tempel.  We
also thank H.~Campins and Y.~R. Feran\'andez for reviewing an early
version of the manuscript, and thank T.~Ootsubo and an anonymous
referee, whose comments improved this paper.


\onecolumn
\clearpage

\clearpage
\setlength\LTcapwidth\textwidth
\renewcommand{\baselinestretch}{0.75}
\renewcommand{\thefootnote}{\alph{footnote}}
\begin{small}
\begin{longtable}{lccccp{2in}}
\caption{Mid-infrared spectroscopy of Jupiter-family comets.\label{table:obs}}
\\[-0.75em]
\hline
\hline
\\[-0.75em]
Comet &
Investigators &
Telescope &
Instrument &
10~\micron{} &
Notes \\
 &
 &
 &
 &
Silicate Band &
 \\
 &
 &
 &
 &
Strength &
\\
\hline
\\[-0.75em]\endfirsthead
\caption[]{\emph{continued}}\\
\\[-0.75em]
\hline
\hline
\\[-0.75em]
Comet &
Investigators &
Telescope &
Instrument &
10~\micron{} &
Notes \\
 &
 &
 &
 &
Silicate Band &
 \\
 &
 &
 &
 &
Strength &
\\
\hline
\\[-0.75em] \endhead
\\[-0.75em]
\hline
\\[-0.75em]
\multicolumn{6}{l}{{\emph{Continued on next page}}} \\
\endfoot
\\[-0.75em]
\hline\\[-5em]
\endlastfoot

2P/Encke                           & \citealt{kelley06apj}          & \spitzer{} & IRS       & N                       & Silicates are $<31\%$ of the sub-\micron{} dust mass ($3\sigma$). \\
4P/Faye                            & \citealt{hanner96}             & IRTF       & BASS      & 1.25                    & \\
9P/Tempel                          & \citealt{lynch95}              & IRAS       & LRS       & $<1.40$\footnotemark[2] & \\
                                   & \citealt{lisse06}              & \spitzer{} & IRS       & Y                       & We find the silicate band strength to be pre-\di:
                                                                                                                         1.19--1.25 (Fig.~\ref{fig:9p-spitzer}); 0.64~hr post-\di: 1.95--2.21\\
                                                                                                                       
                                   & \citealt{harker07}             & Gemini-N   & Michelle  & Y                       & We find the silicate band strength to be pre-\di:
                                                                                                                         May 2005 $1.50\pm0.18$, July 2005 $1.2\pm0.2$;
                                                                                                                         74~min post-\di: 2.02--2.16\\
                                   & \citealt{ootsubo07b}           & Subaru     & COMICS    & Y                       & \\
10P/Tempel                         & \citealt{lynch95}              & IRAS       & LRS       & $<1.40$\footnotemark[2] & \\
19P/Borrelly                       & \citealt{hanner96}             & IRTF       & BASS      & 1.11\footnotemark[2]    & 1.13--1.19 after nucleus subtraction. \\
                                   & \citealt{woodward02}           & IRTF       & HIFOGS    & N                       & \\
24P/Schaumasse                     & \citealt{hanner96}             & IRTF       & BASS      & $<1.10$\footnotemark[2] & Bare nucleus? \\
26P/Grigg-Skjellerup               & \citealt{hanner84}             & UKIRT      & UCL Spec. & N                       & \\
29P/Schwassmann-Wachmann\footnotemark[1] & \citealt{stansberry04}   & \spitzer{} & IRS       & 1.10\footnotemark[2]    & 1.19 after nucleus subtraction. \\
67P/Churyumov-Gerasimenko          & \citealt{hanner85}             & IRTF       & FOGS      & N                       & \\
                                   & \citealt{kelley06apj}          & \spitzer{} & IRS       & \dots                   & \\
69P/Taylor                         & \citealt{sitko04}              & IRTF       & BASS      & 1.23\footnotemark[2]    & 1.24--1.28 after nucleus subtraction. \\
73P/Schwassmann-Wachmann           & \citealt{harker06a}            & Gemini-N   & Michelle  & 1.15--1.25              & Fragments B and C \\
                                   & \citealt{sitko06bass1, sitko06bass2} & IRTF & BASS      & 1.18--1.25              & Fragments B and C \\
                                   & \citealt{sitko06irs-iauc}      & \spitzer{} & IRS       & 1.33                    & Fragment B\\
78P/Gehrels                        & \citealt{watanabe05}           & Subaru     & COMICS    & Y                       & \\
103P/Hartley                       & \citealt{lynch98}              & IRTF       & BASS      & 1.15\footnotemark[2]    & 1.22 after nucleus subtraction. \\
                                   & \citealt{colangeli99}          & ISO        & ISOPHOT   & N                       & The uncertainties may be consistent with $\leq1.2$ \\
                                   & \citealt{crovisier00}          & ISO        & ISOCAM    & 1.20\footnotemark[2]    & The nucleus is $\leq1$\% of the flux.

\footnotetext{\footnotemark[1]The Tisserand parameter of 29P is $T_J =
  2.98$, which places the comet in the Jupiter-family class
  \citep{levison96}, yet in long timescale integrations, the comet
  always remains between the orbits of Jupiter and Saturn \citep[see,
    e.g.,][]{carusi85}---a behavior that is similar to Centaurs.
  Since Centaurs and Jupiter-family comets have the same origin in
  today's solar system (the Kuiper Belt and Scattered Disk) their dust
  properties are related, and a comparison between 29P and JFCs is
  valid.}

\footnotetext{\footnotemark[2]These silicate band strengths do not
  take into consideration the flux from the nucleus.  For most
  entries, we list a corrected value in the notes column.}

\end{longtable}
\end{small}
\renewcommand{\thefootnote}{\arabic{footnote}}
\renewcommand{\baselinestretch}{1.7}


\clearpage
\begin{table}
\begin{center}
\renewcommand{\baselinestretch}{0.7}
\begin{small}
\caption{Coma ($F_C$) and nucleus ($F_N$) fluxes at 10~\micron{} for
  our ``median Jupiter-family comet'' model.$^{\rm
    a}$ \label{table:estimate}}
\begin{tabular}{lllllll}
\\[-0.75em]
\hline
\hline
\\[-0.75em]
 &
Phase &
&
$F_C$ \\
$\Delta$ &
angle &
$F_N$ &
$(\theta=0.20\arcsec)^{\rm b}$ &
\multicolumn{3}{c}{$0.8 F_N / (0.8 F_N + F_C)$ $^{\rm c}$} \\
\cline{5-7}
(AU) &
(\degr) &
(mJy) &
(mJy) &
$\theta=0.20$\arcsec &
$\theta=0.60$\arcsec &
$\theta=1.85$\arcsec \\
\\[-0.75em]
\hline
\\[-0.75em]
\multicolumn{7}{c}{$\afrho=110$~cm, $r_h=1.8$~AU, $R_n=1.8$~km} \\
\hline
0.8 &  0 &  158 &   9 & 0.93 & 0.82 & 0.60 \\
1.0 & 25 &   91 &   7 & 0.91 & 0.77 & 0.52 \\
1.2 & 31 &   60 &   6 & 0.89 & 0.73 & 0.46 \\
\hline \\
\multicolumn{7}{c}{$\afrho=430$~cm$^{\rm d}$, $r_h=1.0$~AU, $R_n=1.8$~km} \\
\hline
0.3 & 81 & 1602 & 481 & 0.73 & 0.47 & 0.22 \\
0.6 & 72 &  488 & 240 & 0.62 & 0.35 & 0.15 \\
0.9 & 63 &  259 & 160 & 0.56 & 0.30 & 0.12 \\
\\[-0.75em]
\hline

\end{tabular}
\end{small}
\end{center}

\renewcommand{\thefootnote}{\alph{footnote}}

\footnotemark[1]{See \S\ref{sec:nucleus} for a discussion of the model
  parameters.  Note that $F_C$, $F_N$, and their ratio
  vary with wavelength.}

\footnotemark[2]{The coma flux scales linearly with aperture angular
  radius, $\theta$, for constant isotropic dust ejection
  (Eq.~\ref{eq:grain2}).}

\footnotemark[3]{The aperture angular radius for the $R\approx100$
  mode of three instruments: $\theta=0.20$\arcsec{} corresponds to
  Gemini/Michelle, $\theta=0.60$\arcsec{} to IRTF/MIRSI, and
  $\theta=1.85$\arcsec{} to \spitzer/IRS.  For simplicity, we assume a
  slit throughput of 80\% for point sources (the nucleus).}

\footnotemark[4]{To estimate the \afrho{} at $r_h=1.0$~AU, we scaled
  the \afrho{} at $r_h=1.8$~AU by a factor of $1.8^{2.3}$ as suggested
  by \citet{ahearn95}.}

\renewcommand{\thefootnote}{\arabic{footnote}}

\renewcommand{\baselinestretch}{1.7}
\end{table}

\end{document}